\documentstyle[graphicx,aps,amstex,epsf]{revtex}
\begin{document}  
\draft \preprint{ }
\twocolumn
\title{Skyrmions in a ferromagnetic Bose-Einstein condensate } 
\author{U.  Al Khawaja and H.  T.  C.  Stoof} \address{Institute
for Theoretical Physics, University of Utrecht, Princetonplein 5,3584 CC
Utrecht, The Netherlands} 
\date{\today} 
\maketitle 
\begin{abstract} 
The recently realized multicomponent Bose-Einstein 
condensates \cite{{mayatt},{stamper},{stenger}} provide opportunities to explore the 
rich physics brought about by the spin degrees of freedom.  
For instance, we can study spin waves and phase separation \cite{hall}, 
macroscopic quantum tunneling\cite{stamper2}, Rabi oscillations\cite{matthews}, 
the coupling 
between spin gradients and superfluid flow, squeezed spin states, 
vortices \cite{matthews2} and other topological excitations. Theoretically, 
there have been already some studies of the ground-state properties
 \cite{{ho},{ohmi},{law},{ho2}} 
of these systems and their line-like vortex excitations
 \cite{{ho},{yip},{williams}}. 
In analogy with nuclear physics or the quantum Hall effect, 
we explore here the possibility of observing point-like 
topological excitations or skyrmions \cite{{skyrme},{skyrme2}}. 
These are nontrivial 
spin textures that in principle can exist in a spinor 
Bose-Einstein condensate. In particular, we investigate 
the stability of skyrmions in a fictitious spin-1/2 
condensate of $^{87}$Rb atoms. We find that skyrmions can exist 
in this case only as a metastable state, but with a 
lifetime of the order of, or even longer than, the typical 
lifetime of the condensate itself. In addition to determining 
the size and the lifetime of the skyrmion, we also present its 
spin texture and finally briefly consider its dynamical properties. 
\end{abstract}


An essential feature of a spinor Bose-Einstein condensate is that 
two or more hyperfine states of the atoms in the condensate have 
almost the same energy. As a result this spin degree of freedom 
becomes a relevant dynamical variable, which gives rise to new 
excitations that are not present in the usual single-component 
Bose-Einstein condensates, where the spins are effectively frozen. 
Among these excitations is the skyrmion, which is a topological 
spin texture. Roughly speaking, the skyrmion is a point-like 
object that can be created out of the ground state, in which 
all the spins are aligned, by reversing the average spin in a 
finite region of space. Although topological considerations 
indeed allow for these excitations, the fundamental question 
that needs to be answered is whether such a configuration is 
also energetically stable. A simple physical argument in favor 
of the stability of the skyrmion excitation is that, although 
gradients of the average local spin increase the kinetic energy 
of the condensate, they also result in a reduction of the density 
of the gas and therefore in a reduction of the interaction energy. 
So we may anticipate that stability occurs when these two competing 
factors are of the same order of magnitude. However, when solving 
the appropriate Gross-Pitaevskii equation, we find that in principle 
this reduction of the interaction energy is always insufficient to 
prevent the skyrmion from collapsing to zero size. Fortunately, 
it turns out that for sufficiently small sizes of the skyrmion 
another stability mechanism starts to work. A number of atoms in 
the centre of the skyrmion, which we denote from now on as the 
core atoms, will be trapped by an effective three-dimensional 
potential barrier, {\it i.e.}, a repulsive shell with a finite radius 
that is induced by the gradients in the spin texture of the 
skyrmion itself. As the skyrmion shrinks in size, the barrier 
height of the repulsive shell increases and the radius decreases. 
This leads to a squeezing of the core atoms and thus to an increase 
in their energy, which ultimately stabilizes the skyrmion. 
Of course, this is not an equilibrium state of the condensate 
since the core atoms will tunnel over the barrier and give the 
skyrmion a finite lifetime. Before discussing this in detail, 
we first need to point out that we perform our calculations 
for a uniform condensate. In spite of this, our results are 
also valid for trapped gases, since the size of the skyrmion 
always turns out to be of the order of the correlation length 
of the condensate, which is typically much smaller than the 
size of the condensate. Moreover, in order to produce realistic 
estimates that can possibly be compared with future experiments, 
we discuss only the case of a spin-1/2 $^{87}$Rb condensate, since the 
spin-1 $^{23}$Na condensate that has also been realized experimentally 
has an antiferromagnetic ground state. Considering other spin 
states or other species is in principle straightforward and will 
be presented elsewhere.
We thus consider a uniform condensate of constant density $n$ and 
constant spinor $\zeta_z=(1,0)$, with all spins being oriented along the 
$z$-axis. 
This represents the ground state of the gas. For the skyrmion, 
however, both $n$ and $\zeta$ are position-dependent. A convenient way 
to take the position dependence of $\zeta({\bf r})$ into account is to express 
the spinor as 
$\zeta({\bf r})=\exp{\left\{-2i{\bf \Omega}({\bf r})
\cdot \hat{{\bf S}}\right\}}\zeta_z$, 
where $\hat {\bf S}$ are the usual angular momentum operators 
for spin-1/2. The physical meaning of this formula is that the 
average spin at a position $\bf r$ is rotated by an angle $2\Omega({\bf r})$ 
from its 
initial orientation, which is in the positive $z$-direction, with  
${\bf \Omega}({\bf r})/{\Omega}({\bf r})$ the axis of rotation. 
An explicit form of ${\bf \Omega}({\bf r})$ determines now a specific 
texture of the skyrmion. We consider only the most symmetric shape 
of skyrmion, because on general grounds this is expected to have 
the lowest energy. We thus take ${\bf \Omega}({\bf r})
={\omega}(r){\bf r}/r$. The boundary conditions at  
$r=0$ and $r\rightarrow \infty$ that the function 
$\omega(r)$ should satisfy are the following. First, 
at $r\rightarrow \infty$ all spins must be oriented as in the ground state, since 
otherwise it requires an infinite amount of energy to create the skyrmion. 
This implies lim$_{r\rightarrow \infty}\omega(r)=0$. 
Along the $z$-axis the spins are also not rotated by 
our ansatz for ${\bf \Omega}({\bf r})$, 
so in order to have a nonsingular texture of the 
spinor with a nonzero winding number, we must require that 
$\omega(0)=2\pi$. 
Finally, to avoid a singular behavior of ${\bf \Omega}({\bf r})$ 
itself, we take only 
functions $\omega(r)$ with zero slope at the origin. In summary,  
$\omega(r)$ is therefore a monotonically decreasing function that starts 
from $2\pi$ at the origin and reaches zero when $r\rightarrow \infty$. 
The specific functional 
form of $\omega(r)$ at intermediate distances between zero and infinity is not 
crucial for the stability of the skyrmion. Only the above boundary 
conditions are important for that. 
From a quantum mechanical point of view, the condensate is 
described by a macroscopic wave function, or order parameter, 
$\psi({\bf r})=\sqrt{n({\bf r})}\zeta({\bf r})$. 
Furthermore, the grand-canonical energy of the gas can in the 
usual mean-field approximation be expressed as a functional 
of $n({\bf r})$ and $\zeta({\bf r})$ \cite{ho}. In detail we have                              
\begin{eqnarray}
E[n({\bf r}),\zeta({\bf r})]&\equiv&\int d({\bf r})\left[
{\hbar^2\over 2m}\left(\nabla{\sqrt{n({\bf r})}}\right)^2
-\mu n({\bf r})\right.\\\nonumber
&+&\left.{\hbar^2\over 2m}n({\bf r})|\nabla \zeta ({\bf r})|^2
+{1\over2}T^{2B}n^2({\bf r})\right]
\label{functional},
\end{eqnarray}                                                                                     
where $m$ is the mass of the atoms, $\mu$ is their chemical potential and  
$T^{2B}=4\pi\hbar^2/2m$ 
is the appropriate coupling constant that represents the strength of 
the inter-atomic interactions in terms of the positive scattering 
length $a$. It is clear from the above expression that the gradients 
in the spin texture lead to an energy contribution, which is 
proportional to $|\nabla \zeta ({\bf r})|^2$. 
Using the above-mentioned form of $\zeta({\bf r})$ and inserting 
the explicit forms of the spin-1/2 matrices, the square of the 
spinor gradient can be written explicitly as  
\begin{equation}
\label{omega}
|\nabla \zeta ({\bf r})|^2=2\left({\sin{(\omega({\bf r}))\over r}}\right)^2
+\left({d \omega\over dr}\right)^2.
\end{equation}                                                                            
In principle, both $n(r)$ and $\omega(r)$ 
can be calculated exactly by minimizing 
the energy functional in Eq. (\ref{functional}) with respect to arbitrary 
functions $n(r)$ and $\omega(r)$ that satisfy the boundary conditions. However, 
the resulting nonlinear and coupled equations are quite difficult 
to solve. Therefore, we use here a variational approach and take for  
the simple ansatz $\omega(r)=4cot^{-2}\left[(r/\lambda)^2\right]$, 
where the variational parameter $\lambda$ physically 
corresponds to the size of the skyrmion. This ansatz is chosen 
here because it automatically incorporates the correct boundary 
conditions and leads to a minimal skyrmion energy as compared 
to various other ansatzes that we have considered. Having specified 
$\omega(r)$, 
we then calculate $n(r)$ exactly by solving numerically the differential 
equation for $n(r)$ obtained by varying $E[n({\bf r}),\zeta({\bf r})]$ 
with respect to $n({\bf r})$. Substituting 
this density profile back into the energy functional, the energy of 
the skyrmion becomes a function of $\lambda$ only and the equilibrium properties 
can be obtained by minimizing this energy with respect to $\lambda$.   
Inserting our ansatz for $\omega(r)$ in Eq. (\ref{omega}), 
$\hbar^2|\nabla\zeta({\bf r})|^2/2m$ takes the shape of an 
off-centered potential barrier with a maximum of $24.3\hbar^2/2m\lambda^2$ 
located at $0.68\lambda$. 
We now distinguish between two cases. The first case occurs when 
the height of the barrier is lower than the chemical potential  
$\mu$ of the atoms. In this case, atoms can move freely across the 
barrier. In the second case, the barrier height is higher than 
the chemical potential and thus atoms become trapped behind the 
barrier near the centre of the skyrmion. Equating 
$\hbar^2|\nabla\zeta({\bf r})|^2/2m$ and $\mu$ gives 
therefore the maximum value of $\lambda_{max}\approx 5\xi$ 
below which trapping takes place. 
Here $\xi$ is the correlation length given by $\xi=1/\sqrt{8\pi an}$, 
and n is the density at $r\gg\lambda$. 
It should be noted that to obtain this value for $\lambda_{max}$, 
we used the fact 
that at sufficiently large radial distances, gradient terms in 
the equation of motion derived from the energy functional vanish, 
and we obtain $\mu=T^{2B}n$. For $\lambda<\lambda_{max}$ 
we calculate first the equilibrium size of the 
skyrmion by minimizing for a fixed number of core atoms the total 
energy of the condensate. Next we also calculate the tunneling 
rate for the core atoms to escape to the outer region (see Appendix). 
The results of these calculations are presented in Fig. 1, where we 
plot the equilibrium size of the skyrmion as a function of the number 
of the core atoms trapped by the texture barrier, together with the 
corresponding tunneling rate. As mentioned previously, we use the 
parameters of $^{87}$Rb with a density of $10^{13}$ cm$^{-3}$. 
We see from this figure 
that only a few atoms in the core of the skyrmion are needed to 
stabilize it and to give it a sufficiently long lifetime. We note 
that the tunneling rate is considerably less than the decay rate of 
the condensate, which is due to two-body collisions and equal to
 \cite{julienne} $Gn$   
with a measured value of \cite{mayatt} $G\approx 2.2\times 10^{-14}$cm$^{-3}$/s. 
We also mention that the lifetime 
becomes even much larger for slightly smaller values of $n$.
Finally, we present some of the interesting properties of the skyrmion, 
which are its texture and its dynamics. The texture can be best presented 
in terms of the three average spin components 
$\left<S_x\right>({\bf r})\equiv 
{\dagger\zeta}({\bf r}){\hat S_x}\zeta({\bf r})$,
$\left<S_y\right>({\bf r})$, and $\left<S_z\right>({\bf r})$ in the three Cartesian 
planes. These are shown in Fig. 2. The most interesting dynamical property of 
the skyrmion comes from the fact that if all spins of the texture are rotated 
around the $z$-axis by a constant angle $\theta$, {\it i.e.},
$\zeta({\bf r})\rightarrow \exp{(-i\theta S_z)}\zeta({\bf r})$, the resulting skyrmion will 
have the same energy. As a result the angle $\theta$ undergoes phase diffusion. 
It also leads to Josephson-like coupling between two skyrmions, which 
will have important consequences for the physics of a skyrmion lattice 
\cite{cote}. 
Another dynamical property is the center of mass motion of the skyrmion, 
which can be shown to be identical to that of a free particle. 
Both dynamical properties will be discussed in detail in a future publication 
(U. and H. T. C., in preparation).

\section*{Appendix}
To calculate the energy of the skyrmion we need to solve the equation for 
the density profile that is obtained from minimizing the energy functional 
in Eq. (\ref{functional}). We solve this equation numerically for the region outside the core. 
Inside the core we solve the equation analytically by using the Thomas-Fermi 
approximation, which amounts to neglecting the gradients of the density profile. 
Using our ansatz for $\omega(r)$, the texture gradient potential 
$V(r)=\hbar^2|\nabla \zeta({\bf r})|^2/2m$ reads 
\begin{equation}
V(r)={32\hbar^2\over 2m\lambda^2}{(r/\lambda)^2\left[3+2(r/\lambda)^4+
3(r/\lambda)^8\right]\over\left[1+(r/\lambda)^4\right]^4}
\label{potential}.
\end{equation}                            
For small $r/\lambda$ this potential can be approximated by a harmonic potential 
with a characteristic frequency $\omega_0 =\sqrt{96\hbar^2/2m^2\lambda^4}$ 
and width $l=\lambda/\sqrt{96}$, as shown in the dotted 
curve in Fig. 3. The use of a Thomas-Fermi approximation is justified 
when the mean-field interaction energy is larger than the spacing 
between the lowest energy levels of the harmonic trap. Specifically, 
the ratio $2Na/l=2\sqrt{96}Na/\lambda$ should be bigger than 1. 
From Fig. 1, we observe that this 
ratio equals approximately 1 for $N=4$ and it increases for larger $N$. 
The lifetime of the skyrmion is estimated by calculating the tunneling 
rate from the core to the outer region over the barrier $V(r)$. To this end 
we employ the following WKB expression for the tunneling rate 
\cite{stoof} 
\begin{equation}
\Gamma\approx{\omega_0\over 2\pi}
\exp{\left[-2\int_{r_1}^{r_2}dr\sqrt{{2m\over\hbar^2}\left(V(r)-
\mu_{core}\right)}\right]}
\label{tunneling},
\end{equation}                                                                 
where $\mu_{core}$ is the chemical potential of the core atoms. 
The radial points $r_1$  
and $r_2$ are the points where $v(r)$ and $\mu_{core}$ 
intersect as shown in Fig. 3. 
The chemical potential $\mu_{core}$ is calculated by differentiating the 
total energy of the core with respect to the number of core atoms.

\section*{Acknowledgements}
The authors would like to thank Michiel Bijlsma for help in the 
numerical calculations and for the helpful remarks. We would also 
like to thank James Anglin, Gerard `t Hooft, David Olive, and Jan 
Smit for useful discussions. This work is supported by the Stichting 
voor Fundamenteel Onderzoek der Materie (FOM), which is financially 
supported by the Nederlandse Organisatie voor Wetenschappelijk Onderzoek (NWO).

\newpage

\section*{Figure Captions}
\begin{figure}[hup]
\begin{center}
\end{center}
\caption{
Decay rate and size of the skyrmion as a function of the 
number of atoms trapped in the core. The two curves are calculated 
for the parameters of a $^{87}$Rb spinor condensate with a density of 
$10^{13}$cm$^{-3}$. 
The size is calculated in units of the correlation length , which 
for these parameters equals $0.85 {\it \mu}$m.
}
\label{fig1}
\end{figure}


\begin{figure}[hup]
\begin{center}
\end{center}
\caption
{
The average spin texture of the skyrmion. Shown are the 
components $\left<S_z ({\bf r})\right>$ and $\left<S_y ({\bf r})\right>$ 
in the three Cartesian planes. A value of $\lambda\approx\xi$, 
which corresponds to 20 core atoms, was chosen to compute these 
quantities. The distances are in units of the coherence length $\xi$. 
The $\left<S_x ({\bf r})\right>$ component can be obtained from the 
$\left<S_y ({\bf r})\right>$ component, by using the 
cylindrical symmetry of the skyrmion. Fig. 1a shows most clearly 
that our skyrmion is composed of two coaxial tori. The two pink 
rings represent the cores of these two tori where 
$\left<S_z ({\bf r})\right>$ reaches its 
maximum negative value -1/2. Fig. 1b shows the cross sections 
of the two tori. 
}
\label{fig2}
\end{figure}
\begin{figure}[hupp]
\begin{center}
\end{center}
\caption{
Figure 3 The potential barrier produced by the skyrmion texture. 
The value of $\lambda$ used for this figure is again approximately $xi$, 
which corresponds to 20 core atoms. The lifetime of the skyrmion is 
determined by the tunneling of core atoms with a chemical potential 
$\mu_{core}$  
to the right of the barrier. The square root of the shaded area is 
the integrand of Eq. (\ref{potential}). The dotted curve represents the harmonic 
approximation to $V(r)$ for small $r/\lambda$. The inset shows 
$\omega(r)$ for the same value 
of $\lambda$.   
 }
\label{fig3}
\end{figure}

\newpage




\end{document}